\documentclass{aa}  

\usepackage{graphicx}
\usepackage{txfonts}
\usepackage{verbatim}
\usepackage{xcolor}

\newcommand{\hii}{\mbox{H\,{\sc ii}}}
\newcommand{\mgii}{\mbox{Mg\,{\sc ii}}}

\newcommand{\ha}{H\,$\alpha$}
\newcommand{\hb}{H\,$\beta$}
\newcommand{\oii}{[O\,{\sc ii}]}
\newcommand{\oiii}{[O\,{\sc iii}]}
\newcommand{\nii}{[N\,{\sc ii}]}
\newcommand{\sii}{[S\,{\sc ii}]}

\newcommand{\mstar}{$\rm M_*$}

\newcommand{\kms}{km\,s$^{-1}$}

\newcommand{\cc}{cm$^{-3}$}

\newcommand{\ergscmarc}{$\rm erg\,s^{-1}\,cm^{-2}\,arcsec^{-2}$}
\newcommand{\msun}{$\rm M_\odot$}

\begin{document}

   \title{Metal line emission around $z<1$ galaxies}

   \author{Rajeshwari Dutta,
          \inst{1}
          Michele Fumagalli\inst{2,3},
          Matteo Fossati\inst{2,4},
          Marc Rafelski\inst{5,6},
          Mitchell Revalski\inst{5},
          Fabrizio Arrigoni Battaia\inst{7},
          Valentina D'Odorico\inst{3,8,9},
          Celine P\'eroux\inst{10,11},
          Laura J. Prichard\inst{5},
          A. M. Swinbank\inst{12}
          }

   \institute{IUCAA, Postbag 4, Ganeshkind, Pune 411007, India \newline
              \email{rajeshwari.dutta@iucaa.in}
         \and     
              Dipartimento di Fisica G. Occhialini, Universit\`a degli    Studi di Milano Bicocca, Piazza della Scienza 3, 20126 Milano, Italy
         \and     
              INAF – Osservatorio Astronomico di Trieste, via G. B. Tiepolo 11, I-34143 Trieste, Italy
         \and     
              INAF - Osservatorio Astronomico di Brera, via Bianchi 46, 23087 Merate (LC), Italy
         \and
              Space Telescope Science Institute, 3700 San Martin Drive, Baltimore, MD 21218, USA
         \and
              Department of Physics and Astronomy, Johns Hopkins University, Baltimore, MD 21218, USA
         \and
              Max-Planck-Institut f\"ur Astrophysik, Karl-Schwarzschild-Str. 1, D-85748 Garching bei M\"unchen, Germany
         \and
              Scuola Normale Superiore, P.zza dei Cavalieri, I-56126 Pisa, Italy
         \and
              IFPU-Institute for Fundamental Physics of the Universe, via Beirut 2, I-34151 Trieste, Italy
         \and 
              European Southern Observatory, Karl-Schwarzschildstrasse 2, D-85748 Garching bei M\"unchen, Germany
         \and
              Aix Marseille Universit\'e, CNRS, LAM (Laboratoire d'Astrophysique de Marseille) UMR 7326, F-13388, Marseille, France
         \and
              Centre for Extragalactic Astronomy, Department of Physics, Durham University, South Road, Durham DH1 3LE, UK
             }

    \authorrunning{R. Dutta et al.}

   \date{Received ; accepted }

\abstract{We characterize, for the first time, the average extended emission in multiple lines (\oii, \oiii, and \hb) around a statistical sample of 560 galaxies at $z\approx0.25-0.85$. By stacking the Multi Unit Spectroscopic Explorer (MUSE) 3D data from two large surveys, the MUSE Analysis of Gas around Galaxies (MAGG) and the MUSE Ultra Deep Field (MUDF), we detect significant \oii\ emission out to $\approx$40 kpc, while \oiii\ and \hb\ emission is detected out to $\approx$30 kpc. Via comparisons with the nearby average stellar continuum emission, we find that the line emission at 20--30 kpc likely arises from the disk-halo interface. Combining our results with that of our previous study at $z\approx1$, we find that the average \oii\ surface brightness increases independently with redshift over $z\approx$0.4--1.3 and with stellar mass over \mstar\ $\approx10^{6-12}$\,\msun, which is likely driven by the star formation rate as well as the physical conditions of the gas. By comparing the observed line fluxes with photoionization models, we find that the ionization parameter declines with distance, going from log q (cm~s$^{-1}$) $\approx$7.7 at $\le$5 kpc to $\approx$7.3 at 20--30 kpc, which reflects a weaker radiation field in the outer regions of galaxies. The gas-phase metallicity shows no significant variation over 30 kpc, with a metallicity gradient of $\approx$0.003 dex kpc$^{-1}$, which indicates an efficient mixing of metals on these scales. Alternatively, there could be a significant contribution from shocks and diffuse ionized gas to the line emission in the outer regions.}


   \maketitle
%

\defcitealias{Dutta2023}{D23}

\section{Introduction} 
\label{sec:intro}

The majority of the baryons in the Universe are found in the form of gas in the intergalactic medium (IGM) and the circumgalactic medium (CGM) of galaxies. The gas acts as the fuel for the formation of stars in galaxies. The formation of stars leads to the creation of heavy elements or metals. Stellar winds and supernova explosions eject the metals out of stars into the interstellar medium (ISM) and even beyond into the CGM and the IGM. Galaxies thus form and evolve in complex ecosystems that are regulated by a cycle of gas flows \citep{Somerville2015,Tumlinson2017,Peroux2020a}. This baryon cycle is believed to consist of: metal-poor gas accretion onto galaxies from filaments of the IGM; stellar feedback and active galactic nucleus-driven outflows of metal-enriched gas into the CGM; recycling of gas from the CGM back into the ISM through fountains; and transfer of gas between galaxies through interactions and mergers \citep{Oppenheimer2008,Dekel2009,Cresci2010,Kewley2010,Lilly2013,Fraternali2017}. All these processes leave their imprint on the distribution and physical conditions of gas within and around galaxies, with the relative importance of each process evolving with cosmic time.

Strong rest-frame optical emission lines have long been used to trace the physical conditions of the ionized gas in the ISM. Several diagnostic line ratio diagrams have been developed to probe the chemical and ionization state of the ISM across redshifts \citep[e.g.,][]{Kauffmann2003,Kewley2019,Maiolino2019}. Higher-redshift galaxies are found to have higher ionization parameters, harder ionizing spectra, and higher electron densities \citep[e.g.,][]{Nakajima2013,Shirazi2014,Sanders2021}. Furthermore, the metallicity of the gas in the ISM is found to decrease with increasing redshift at a fixed stellar mass \citep[e.g.,][]{Erb2006,Maiolino2008,Sanders2021}. 

In the nearby Universe, the gas-phase metallicity in galaxies typically exhibits a negative gradient (i.e., the central regions are more metal-rich than the outskirts), which supports an inside-out picture of galaxy evolution \citep[e.g.,][]{Zaritsky1994,Sanchez2014,Belfiore2017}. Integral field unit (IFU) spectrographs, such as the Multi Unit Spectroscopic Explorer \citep[MUSE;][]{Bacon2010} and the K-band Multi Object Spectrograph \citep[K-MOS;][]{Sharples2013} on the Very Large Telescope (VLT), and slitless grism spectrographs on the \textit{Hubble} Space Telescope (HST) have enabled spatially resolved studies of the ISM up to $z\approx3$. Such studies have revealed flatter metallicity gradients in $z\gtrsim0.5$ galaxies with a large scatter and some galaxies even having positive gradients \citep[e.g.,][]{Wuyts2016,Carton2018,Curti2020,Wang2020,Simons2021}. Recently, \textit{James Webb }Space Telescope (JWST) near-infrared (NIR) IFU observations found flat metallicity gradients in $z\approx6-8$ galaxies \citep{Venturi2024}. The flattening of metallicity gradients could be due to several factors, including the inflow of metal-poor gas into the inner regions, outflows populating the outer regions with metals, recycling of metal-rich gas in the outer regions, and transfer of metals due to galaxy interactions and mergers \citep[e.g.,][]{Cresci2010,Kewley2010,Rupke2010,Bresolin2012}.

The above-cited studies probed the ionized gas within the stellar disks of galaxies (a few kiloparsecs, or a few effective radii). Beyond the stellar disks, galaxies are surrounded by halos of diffuse gas that extend out to a few hundred kiloparsecs (or a few virial radii). This is known as the CGM \citep{Tumlinson2017}, and it is relatively challenging to probe in emission due to the low gas density but is effectively probed in absorption against a bright background source such as a quasar. Along with the stellar and ISM properties of galaxies, the distribution and physical conditions of gas in the CGM can place crucial constraints on models of galaxy formation and evolution \citep[e.g.,][]{Fumagalli2016,Liang2016,Turner2017,Oppenheimer2018,Weng2024}. Based on several large observational campaigns carried out to characterize the CGM, we know that the CGM  at $z\lesssim4$ consists of metal-enriched, multiphase gas, which becomes more ionized with increasing distance from galaxies and which is influenced by both galaxy properties and environmental effects \citep[e.g.,][]{Werk2014,Hamanowicz2020,Dutta2020,Dutta2021,Galbiati2023,Qu2023}. 

Detecting emission in multiple lines from the diffuse gas in the extended stellar disk, disk-halo interface, and outer halo can provide us with additional and complementary insights into the distribution and physical conditions of the CGM. Sensitive optical IFU spectrographs such as MUSE and the Keck Cosmic Web Imager \citep[KCWI;][]{Morrissey2018} have facilitated detections of the low surface brightness (SB) halo gas at cosmological distances in emission. There have been a few detections of metal line emission extending up to tens of kiloparsecs around both individual galaxies and stacks of galaxies \citep[e.g.,][]{Burchett2021,Zabl2021,Leclercq2022,Leclercq2024,Dutta2023,Guo2023}. 

Recently, by stacking the MUSE data of a sample of $\approx$600 galaxies, \citet{Dutta2023} detected, for the first time, the average \mgii\ and \oii\ line emission up to $\approx$30--40 kpc from a general population of galaxy halos at $z\approx1$. The shallower radial SB profile of \mgii\ up to $\approx$20 kpc compared to \oii\ suggests that the resonant \mgii\ emission is affected by dust and radiative transfer effects, while the constant ratio of \oii\ to \mgii\ over $\approx$20--40 kpc suggests a non-negligible in situ origin of the extended metal emission. At $z<1$, the detection of extended emission in multiple lines (e.g., \ha, \hb, \oii, and \oiii) around a few galaxies and quasars has enabled studies of the physical conditions and excitation mechanisms in the gas beyond the stellar disks \citep[e.g.,][]{Johnson2018,Chen2019,Nielsen2023}. Such studies help bridge our understanding of the gas within the stellar disks and the gas beyond it in the halo. 

Here we extend the stacking analysis of \citet{Dutta2023} to $z<1$ to investigate the average physical conditions in gas around a statistical sample of galaxies. The observations and the stacking procedure used in this work are explained in Sect.~\ref{sec:obs}. The results of the stacked line emission and line ratios around galaxies are presented in Sect.~\ref{sec:results}. The results are summarized and discussed in Sect.~\ref{sec:discussion}. Throughout this work, we use a \textit{Planck} 2015 cosmology with $H_{\rm 0}$ = 67.7\,\kms\,Mpc$^{-1}$ and $\Omega_{\rm M}$ = 0.307 \citep{Planck2016}.

\section{Observations and analysis} 
\label{sec:obs}

\begin{figure}
    \centering
    \includegraphics[width=0.5\textwidth]{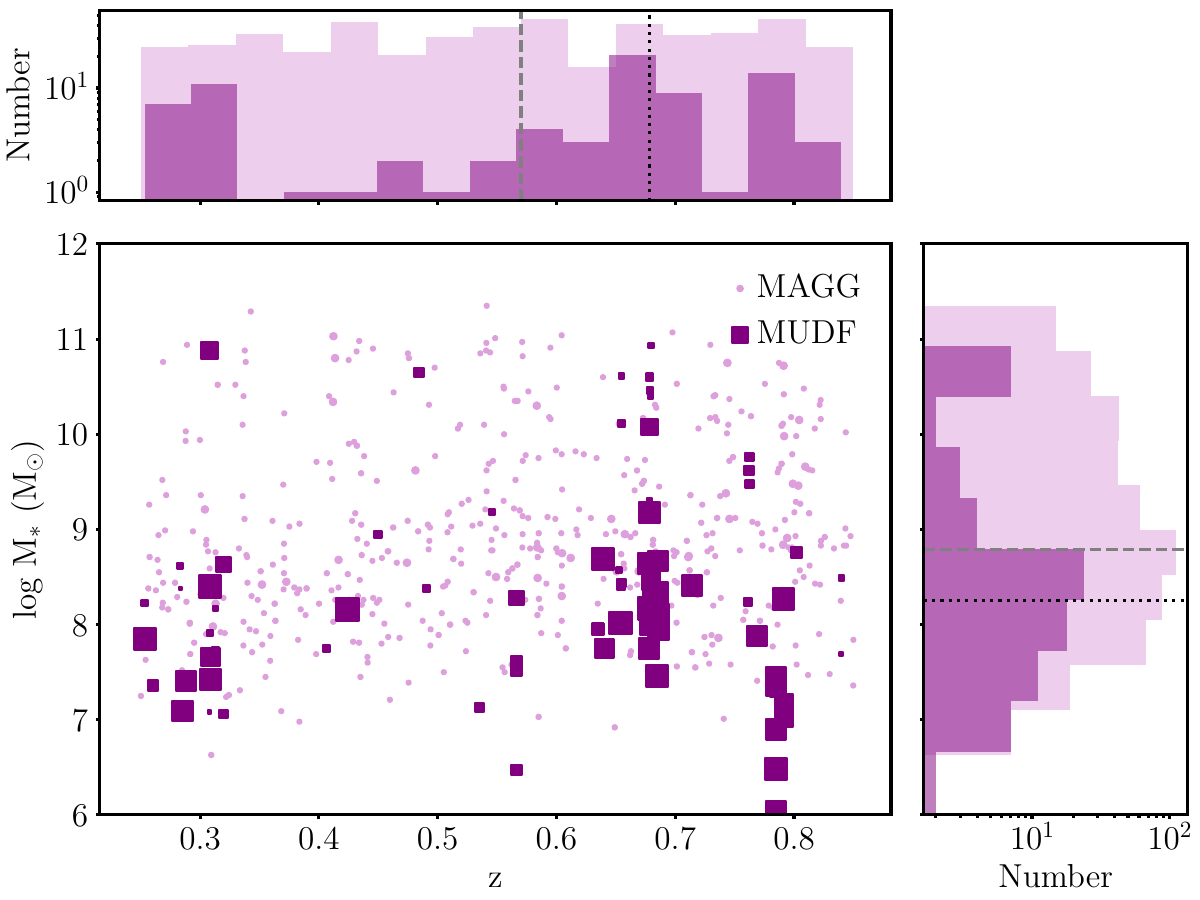}
    \caption{Sellar mass versus redshift of the sample of galaxies used for stacking. The galaxies from the MAGG survey are marked as light purple circles, and the galaxies from the MUDF survey are marked as dark purple squares. The size of the symbols is proportional to the MUSE exposure time of the galaxies. The top and right panels show the histograms of redshifts and stellar masses, respectively, in light purple for the MAGG galaxies, and in dark purple for the MUDF galaxies. The dashed lines mark the median values of the MAGG galaxies, and the dotted lines mark the median values of the MUDF galaxies.}
    \label{fig:mstar_zgal}
\end{figure}

This work is based upon MUSE observations from two large programs on the VLT: (1) MUSE Analysis of Gas around Galaxies \citep[MAGG;][]{Lofthouse2023}, which consists of medium-deep, single pointing (1$\times$1 arcmin$^2$) MUSE observations of 28 fields that are centered on quasars at $z\approx3.2-4.5$; (2) MUSE Ultra Deep Field \citep[MUDF;][]{Fossati2019}, which consists of very deep MUSE observations of a 1.5$\times$1.2 arcmin$^2$ region around a pair of quasars at $z\approx3.2$. The MUSE exposure time in the MAGG survey is $\approx$4 h per field, except for two fields that have deeper exposure time of $\approx$10 h. The MUSE observations in the MUDF survey amount to a total of $\approx$143 h, with the exposure time decreasing from the center of the field to the outer regions. The average image quality of the MUSE data is $\approx$0.7 arcsec full-width at half-maximum (FWHM), which corresponds to $\le$5 kpc at $z<1$. 

The observations and galaxy catalogs from these surveys are described in detail in Sect. 2 of \citet{Dutta2023}. We selected all the galaxies in the MAGG and MUDF catalogs that lie within the redshift range $z=0.25-0.85$. This selection ensured that we got simultaneous coverage of the \oii\ $\lambda\lambda$3727,3729, \hb\ $\lambda$4863, \oiii\ $\lambda$4960 and $\lambda$5008 emission lines in the MUSE spectra. The total sample consists of 560 galaxies, 480 from the MAGG survey, and 80 from the MUDF survey. The total MUSE exposure time of the galaxies used for stacking in this work is $\approx$6264 h.

To obtain the physical properties of the galaxies such as stellar masses and star formation rates (SFRs), the MUSE spectra and photometry were jointly fit with stellar population synthesis models using the Monte Carlo Spectro-Photometric Fitter \citep[{\sc mc-spf};][]{Fossati2018}. For the MUDF galaxy sample, HST optical and NIR photometry in five bands \citep{Revalski2023} and HAWK-I K-band photometry were used in addition to the MUSE spectra and photometry. We refer to Sect. 2 of \citet{Dutta2020} for details of the spectral energy distribution (SED) fitting process. Briefly, {\sc mc-spf} adopts the \citet{Bruzual2003} models at solar metallicity, the \citet{Chabrier2003} initial mass function, nebular emission lines from the models of \citet{Byler2018}, and the dust attenuation law of \citet{Calzetti2000}.

The distributions of redshifts and stellar masses of the galaxy sample used for stacking in this work are shown in Fig.~\ref{fig:mstar_zgal}. The median redshift of the sample is $\approx$0.6. The typical uncertainty in the redshift estimates from MUSE spectra is $\approx$60\,\kms. The galaxy sample spans the stellar mass range of $\approx$$10^{6-11}$\,\msun, with a median stellar mass of $\approx$$10^9$\,\msun. The typical uncertainty in the estimates of stellar mass from SED fitting is $\approx$0.1--0.2 dex.

We followed the same procedure to stack the MUSE data of the galaxies as described in Sect. 3 of \citet{Dutta2023}. In brief, for each galaxy we first extracted a sub-cube of area 200$\times$200 kpc$^2$ centered on the galaxy spanning 100\,\AA\ in rest wavelength around the emission lines of interest, after subtracting the continuum emission and masking out all the other continuum sources. We performed both mean and median stacking of all the sub-cubes in the rest-frame wavelengths of the galaxies. The results from both the stacks are found to be consistent within 1$\sigma$. Here we present the results based on median stacking, which is more robust against outliers. To subtract the continuum, we fit a low-order spline to the continuum emission around the lines for each spaxel in the cube. We repeated the above procedure 100 times with repetition to obtain the sample variance from the 16$^{\rm th}$ and 84$^{\rm th}$ percentiles of the bootstrapped sample.

We corrected for the Milky Way extinction following \citet{Schlafly2011} and the \citet{Fitzpatrick1999} extinction curve. We applied a correction for dust to the fluxes of each galaxy before stacking using the dust extinction derived by {\sc mc-spf} and the extinction curve of \citet{Calzetti2000}. We checked that there is no significant difference in the results if instead we correct for dust after stacking using the median dust extinction of the sample. We note that there likely are radial gradients in the dust attenuation within a galaxy \citep[e.g.,][]{Wuyts2012,Nelson2016,Tacchella2018,Wilman2020}, which we do not take into account in this work because spatially resolved dust attenuation distribution in high-$z$ galaxies is still not well understood, particularly at the larger spatial scales probed in this work. For the median stellar mass of this sample, the dust attenuation is found to be weak and flat on average in the ISM of $z\approx1$ galaxies \citep{Nelson2016}. To correct the \hb\ flux for the underlying stellar absorption, we increased the flux by 3\% based on the typical average correction found in the literature \citep[e.g.,][]{Cullen2021,Sanders2021,Curti2023,Stephenson2024}.

To generate pseudo-narrowband (NB) images of the line emission, we summed the stacked cube from -250 to 250 \kms\ in rest-frame velocity around the \hb\ and \oiii\ lines, and from -250 to 500 \kms\ around the $\lambda$3727 line of the \oii\ doublet to cover both the lines of the doublet. This velocity range is found to encompass all the emission in the stacked spectra (see Fig.~\ref{fig:spectra}). For reference, the median virial velocity expected for the host halos of the galaxies in this sample is $\approx$100\,\kms. To make a comparison with the line NB images, we generated NB images of the stellar continuum emission by averaging the NB images over two velocity windows around $\pm$1500\,\kms\ of each line with the same velocity width as for the line NB. The line and continuum NB images were used to estimate the average SB radial profiles. We did not take the inclination and orientation of the galaxies  into account during stacking because we did not have HST imaging available for the full sample. Therefore, the stacked emission at any projected separation represents the azimuthally averaged emission of the galaxy sample at that separation.

\section{Results} 
\label{sec:results}

\subsection{Line emission around galaxies}
\label{sec:results_lines}

\begin{figure*}
    \centering
    \includegraphics[width=1\textwidth]{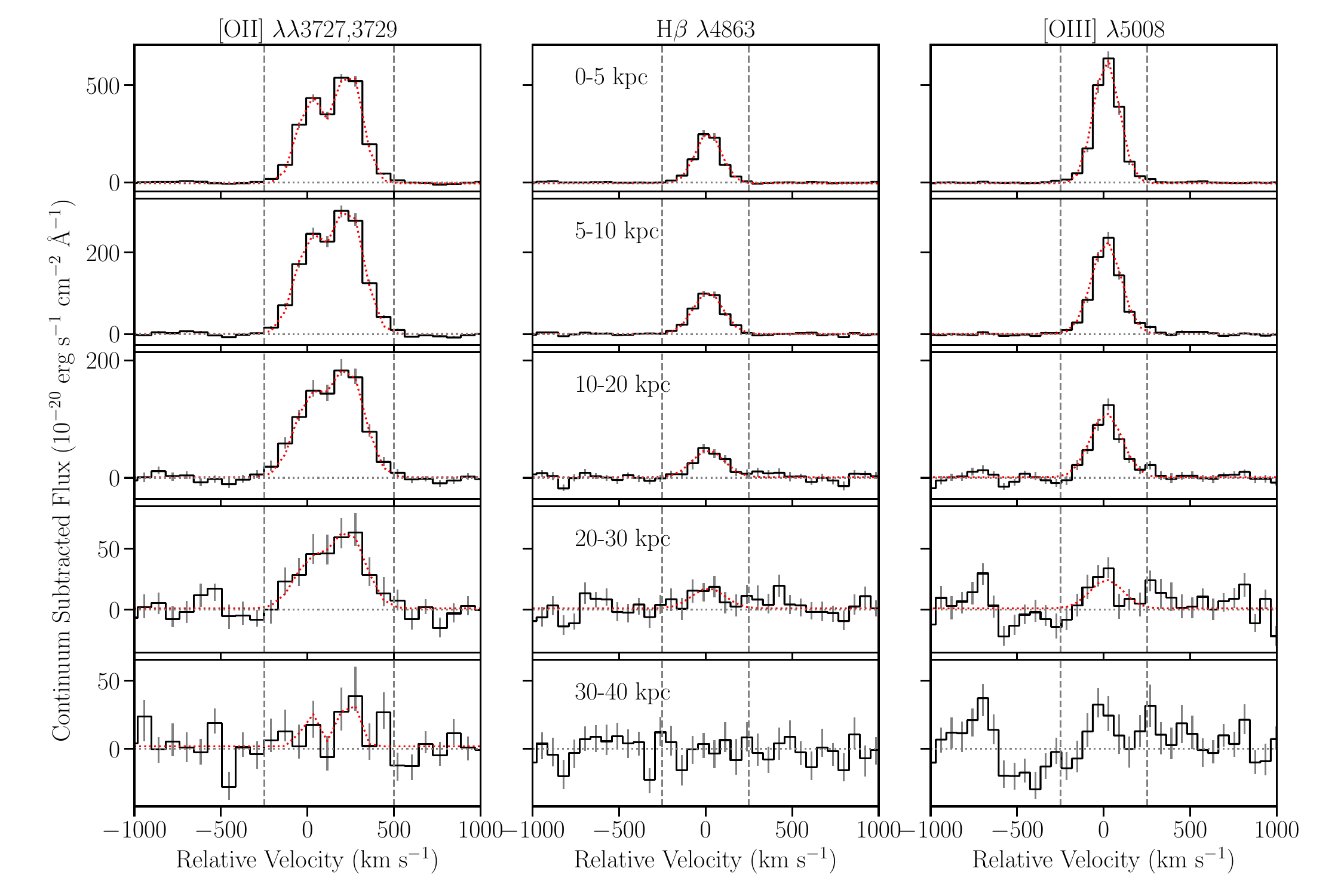}
    \caption{Spectra of different emission lines (labeled at the top of each column) extracted from the median stacked cubes of the sample. From top to bottom, the spectra represent the total flux in annuli in radius ranges of 0--5 kpc, 5--10 kpc, 10--20 kpc, 20--30 kpc, and 30--40 kpc. The 16$^{\rm th}$ and 84$^{\rm th}$ percentiles of the spectra from bootstrapping analysis are shown as vertical bars. The Gaussian fits to the lines are shown as dotted red lines. The velocity windows used to construct the pseudo-NB images and SB profiles are marked with vertical dashed lines.}
    \label{fig:spectra}
\end{figure*}

\begin{figure*}
    \centering
    \includegraphics[width=1\textwidth]{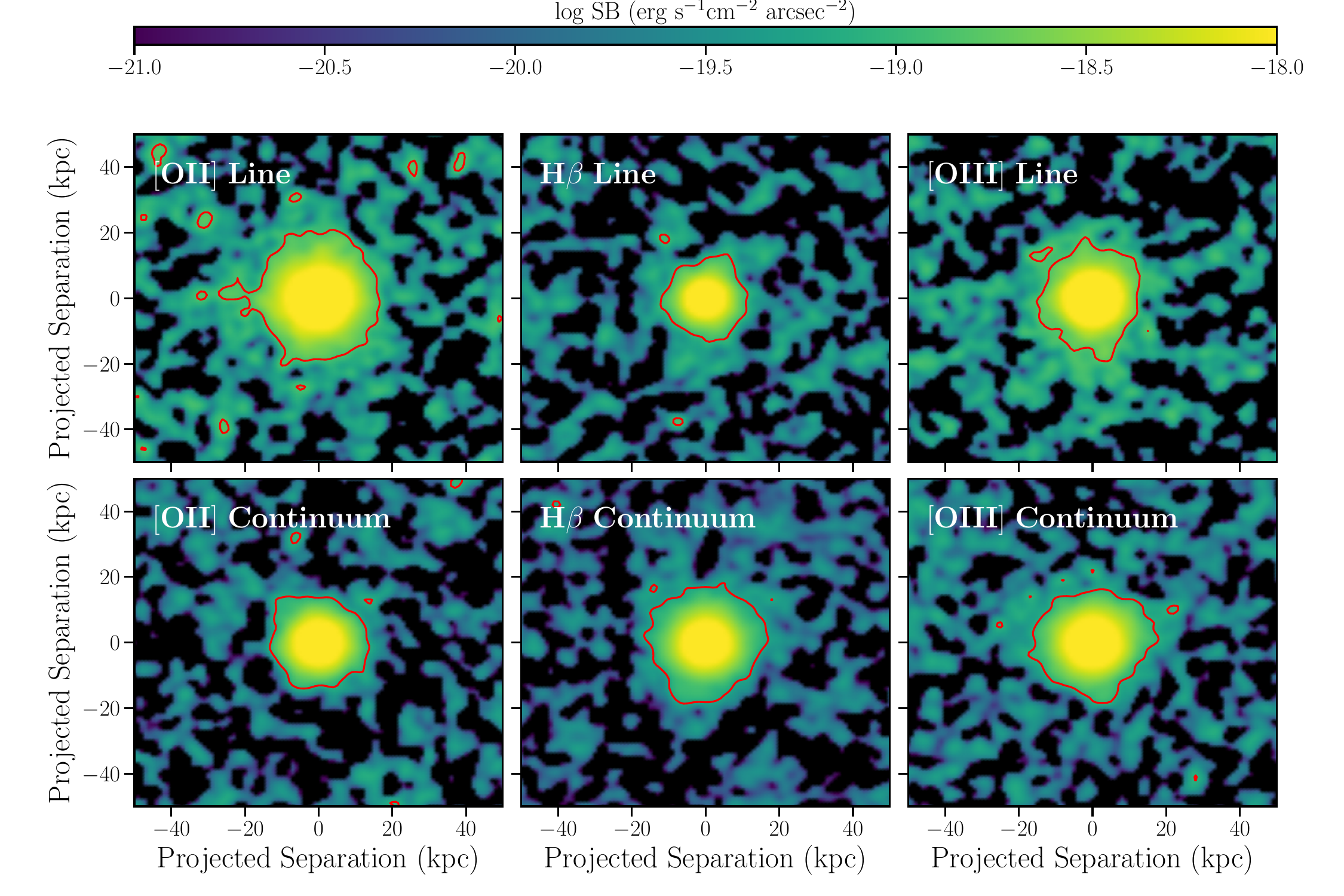}
    \caption{Pseudo-NB images of the median stacked sample smoothed using a Gaussian kernel of standard deviation 0.3 arcsec. The top row shows, from left to right, the NB images of the \oii\ ($\lambda$3727+$\lambda$3729), \hb, and \oiii\ ($\lambda$4960+$\lambda$5008) line emission. The bottom row shows, from left to right, the NB images of the continuum emission near the \oii, \hb, and \oiii\ lines. The red contour is plotted at three times the rms noise in the NB images. }
    \label{fig:nb_images}
\end{figure*}

\begin{figure*}
    \centering
    \includegraphics[width=1\textwidth]{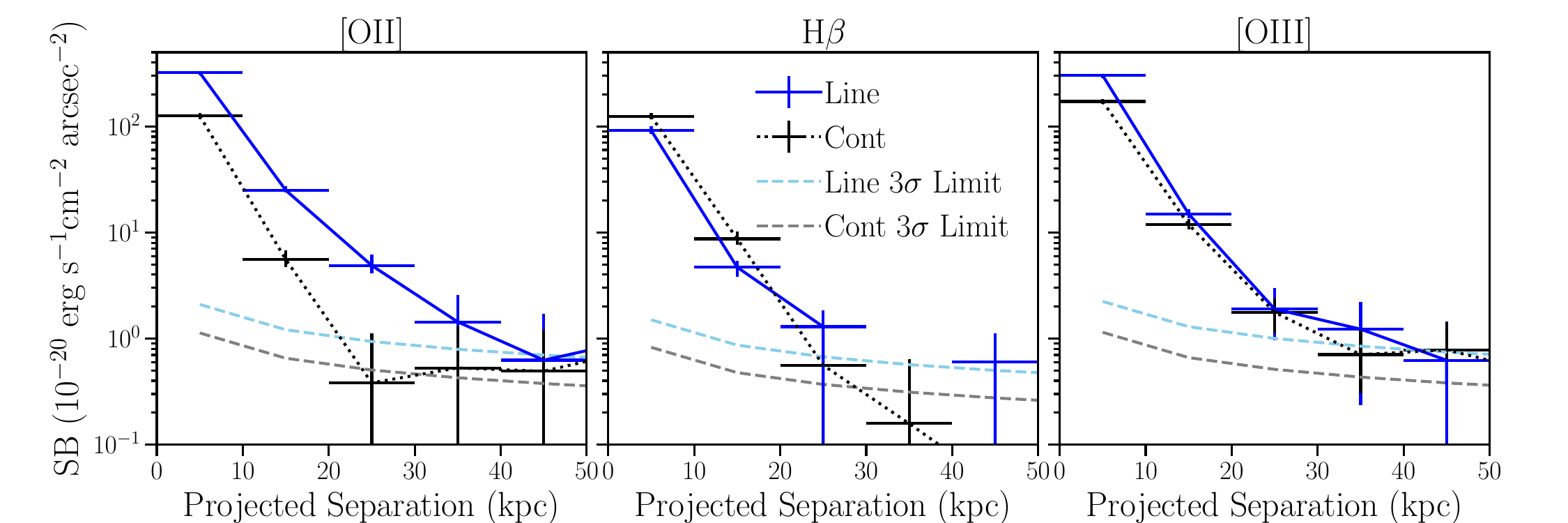}
    \caption{SB of the median stacked sample as a function of projected separation from the galaxy center. The panels from left to right show the SB profiles of the \oii\ ($\lambda$3727+$\lambda$3729), \hb, and \oiii\ ($\lambda$4960+$\lambda$5008) emission. The SB is estimated as the azimuthally averaged value in annuli of radius 10 kpc from the pseudo-NB images shown in Fig.~\ref{fig:nb_images}. In each panel, the solid blue line shows the SB profile of the line emission, the dotted black line shows the SB profile of the continuum emission near the line (averaged over two velocity windows with the same width as that of the line), the dashed light blue line shows the $3\sigma$ upper limit on the line SB, and the dashed gray line shows the $3\sigma$ upper limit on the continuum SB. The 16$^{\rm th}$ and 84$^{\rm th}$ percentiles of the SB profiles from bootstrapping analysis are shown as vertical bars. Note that the average image quality of the MUSE data is $\le$5 kpc FWHM at $z\le1$.}
    \label{fig:sb_profiles}
\end{figure*}

First we characterized the average emission around $z\approx0.6$ galaxies from the \oii, \hb, and \oiii\ lines. We extracted the 1D spectra from the 3D median stacked cubes by summing up the flux over different annular regions. Figure~\ref{fig:spectra} shows the spectra extracted within 0--5 kpc, 5--10 kpc, 10--20 kpc, 20--30 kpc, and 30--40 kpc around the \oii\ $\lambda\lambda$3727,3729 doublet, \hb, and \oiii\ $\lambda$5008 lines. We note that the \oiii\ $\lambda$4960 line, not shown here, is found to have 1/3 the flux of the \oiii\ $\lambda$5008 line as per theoretical expectations. The \oii\ line emission is detected out to $\approx$40 kpc at 3$\sigma$, while the \hb\ and \oiii\ emission lines are detected out to $\approx$30 kpc at 3$\sigma$. It can be seen that, as expected, the flux of the emission lines decreases going outward from the center. We fit Gaussian profiles to the four emission lines with the constraint that the velocity centroid and velocity width of all the lines are equal. We fit a double Gaussian profile to the \oii\ doublet, allowing the ratio of the line fluxes, $\lambda$3729/$\lambda$3727, to vary between the low density limit of 0.35 and the high density limit of 1.5. We find that the line fluxes decrease by a factor of $\approx$10--20 going from the central 5 kpc region to the outer 20--30 kpc annular region. In addition to the decrease in flux, the emission lines become broader in the outer regions, with the velocity dispersion (not corrected for line spread function) going from $\approx$75\,\kms\ within 5 kpc to $\approx$96\,\kms\ at 20--30 kpc. The \oii\ doublet ratio increases from $\approx$1.3 within 5 kpc to the low density limit of 1.5 at 30--40 kpc.

The top row of Fig.~\ref{fig:nb_images} shows the NB images of the line emission in the median stacked cube. The NB images of the continuum emission around the lines are shown in the second row for comparison. The combined NB image of the \oiii\ $\lambda$4960 and $\lambda$5008 lines is shown. The \oii\ line emission is the most spatially extended, followed by the \oiii\ and \hb\ emission. To quantify the average radial profile and extent of the emission, the azimuthally averaged SB in annuli of radius 10 kpc, as obtained from the NB images, are shown in Fig.~\ref{fig:sb_profiles}. The \oii\ SB profile is more radially extended than that of the continuum, while the \hb\ and \oiii\ SB profiles are similar to that of the continuum. The continuum emission near the \oii\ line extends out to $\approx$20 kpc at a 3$\sigma$ SB level of $\approx10^{-20}$\,\ergscmarc. This suggests that we are probing the stellar disk or ISM up to $\approx$20 kpc. Beyond that at 20--30 kpc, the \oii\ emission is significantly enhanced compared to the continuum, while \hb\ and \oiii\ are marginally enhanced. The line emission in this region is most likely originating from the disk-halo interface, where the ISM is transitioning into the CGM. In the case of \oii, we detect significant emission out to 30--40 kpc, which likely originates from the CGM.

\subsection{Evolution of {\rm \oii} emission}
\label{sec:results_oii}

\begin{figure*}
    \centering
    \includegraphics[width=0.45\textwidth]{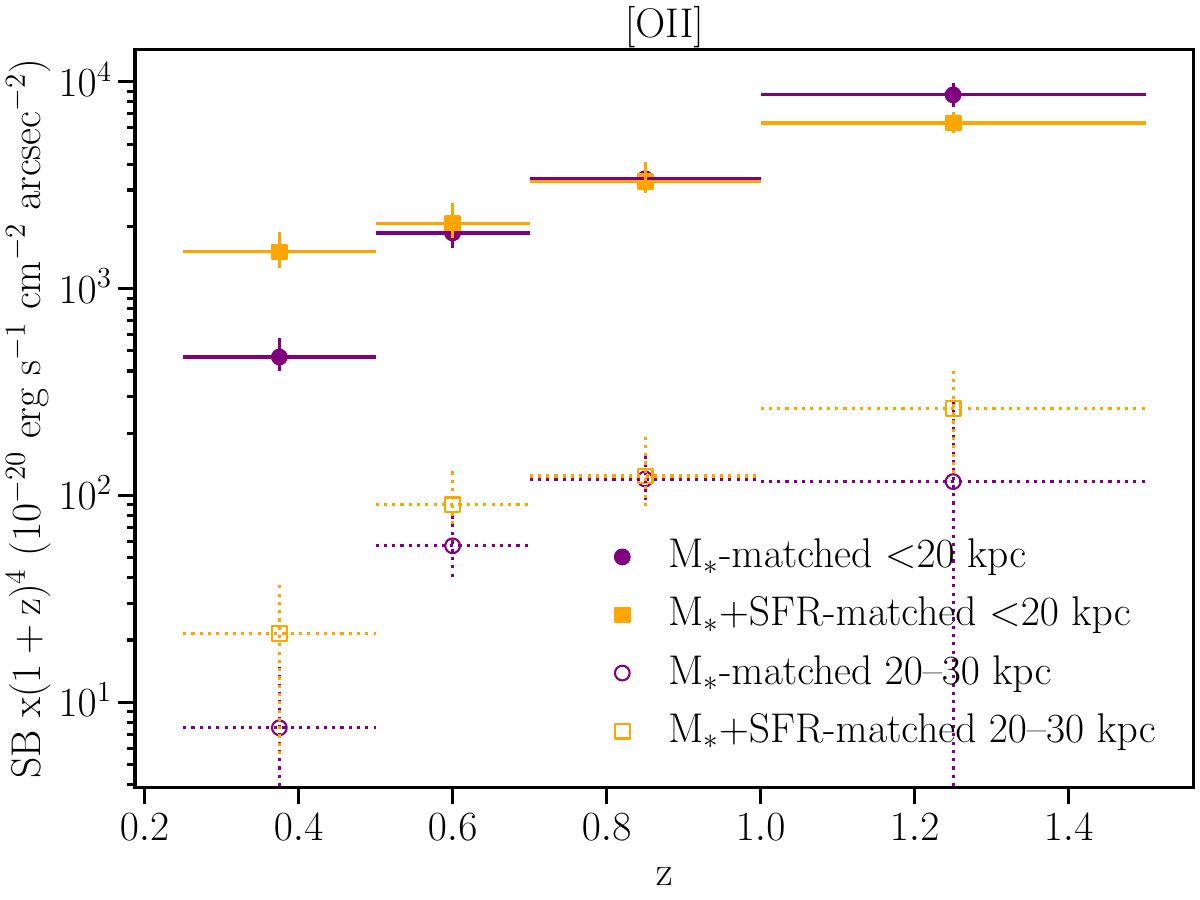}
    \includegraphics[width=0.45\textwidth]{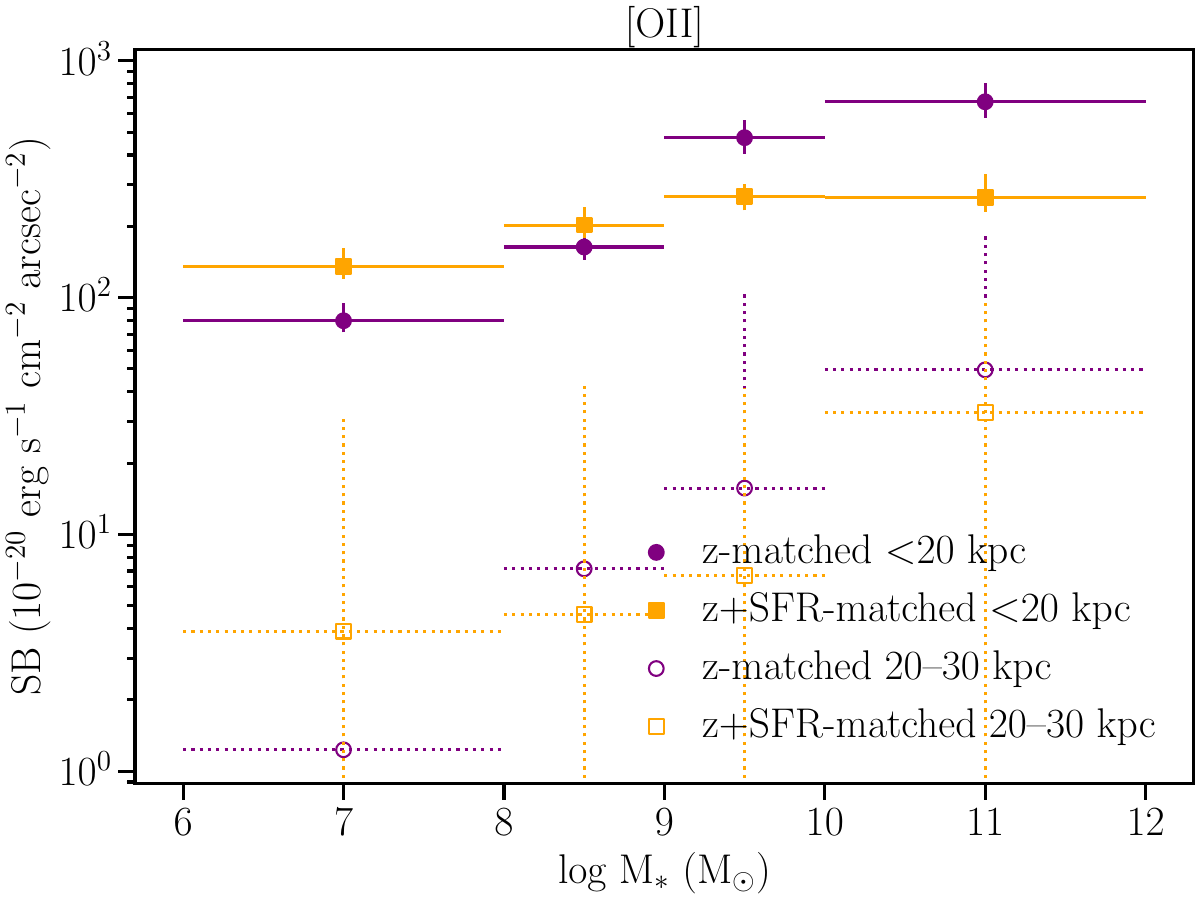}
    \caption{Evolution of \oii\ SB with redshift and stellar mass. Left: Average SB of \oii\ emission, corrected for cosmological dimming, in four different redshift bins. The purple circles are for samples that are matched in stellar mass, whereas the orange squares are for samples that are matched in both stellar mass and SFR. 
    Right: Average SB of \oii\ emission in four different stellar mass bins. The purple circles are for samples that are matched in redshift, whereas the orange squares are for samples that are matched in both redshift and SFR. In both panels, the filled symbols represent the average values within a circular aperture of radius 20 kpc, and the open symbols the average values within an annular region between radii 20 and 30 kpc.}
    \label{fig:sb_oii}
\end{figure*}

Based on a similar stacking analysis of higher-redshift galaxies ($z\approx0.7-1.5$) in the MAGG and MUDF surveys, \citet{Dutta2023} found that the \oii\ line emission becomes brighter and more spatially extended with increasing stellar mass and redshift, and suggested that this was likely due to higher SFRs at higher stellar masses and redshifts. To further investigate the evolution of \oii\ emission with redshift, we combined our sample with that presented in \citet{Dutta2023}, and conducted two control experiments. First, we formed four subsamples in the redshift ranges, $0.25 \le z < 0.5$, $0.5 \le z < 0.7$, $0.7 \le z < 1.0$, and $1.0 \le z < 1.5$, which are matched in stellar mass within a factor of two, such that the maximum difference between the cumulative distributions of the stellar mass in each of the redshift bins is $\lesssim$0.1 and the $p$-value is $\gtrsim$0.8 based on two-sided Kolmogorov-Smirnov test. In this way, we were able to form a stellar mass-matched sample of 70 galaxies in each of the four redshift bins. Next, similar to above, we formed subsamples in the four redshift bins that are matched within a factor of two in both stellar mass and SFR. This led to a matched sample of 34 galaxies in each redshift bin. 

In the left panel of Fig.~\ref{fig:sb_oii} we plot the average \oii\ SB, corrected for cosmological dimming, as a function of redshift, for both the matched samples. The average values are estimated in two regions: within a circular aperture of radius 20 kpc that that probes the ISM and within an annular aperture between radii 20 and 30 kpc that likely probes the disk-halo interface. For the inner region, the average SB is found to increase by a factor of $\approx$20 from $z\approx0.4$ to $z\approx1.3$ in the case of the stellar mass-matched samples, while in the case of the stellar mass- and SFR-matched samples, the dependence of SB on redshift is weaker, with the average SB increasing by a factor of $\approx$4 from $z\approx0.4$ to $z\approx1.3$. For the outer region, the trend with redshift again becomes weaker for the stellar mass- and SFR-matched samples, although there is no significant difference between the two matched samples. This indicates that the higher SFR of galaxies at higher redshifts is likely a major factor behind the enhanced \oii\ emission in the ISM at high redshifts. However, there are likely additional factors, such as gas density, that contribute to the enhanced extended \oii\ emission at high redshifts. Indeed, studies have found that the density of the warm ionized gas is higher at $z\approx2-3$ compared to local galaxies that are matched in stellar mass and star formation activities \citep[e.g.,][]{Shirazi2014,Davies2021}.

Furthermore, we performed similar control experiments as above to investigate the dependence of \oii\ emission on stellar mass using the combined sample of this work and \citet{Dutta2023}. Firstly, we formed four subsamples in the stellar mass bins, \mstar\ = $10^{6-8}, 10^{8-9}, 10^{9-10}$, and $10^{10-12}$\,\msun, which are matched in redshift within 0.3, such that the $p$-value from Kolmogorov-Smirnov test is $\gtrsim$0.8. This led to a redshift matched sample of 51 galaxies in each of the stellar mass bins. Secondly, we formed four subsamples in the above stellar mass bins that are matched within a factor of two in SFR in addition to being matched in redshift as above, leading to samples of 25 galaxies in each mass bin. We then performed the \oii\ line emission stacking for these subsamples. The average \oii\ SB values in the inner 20 kpc and outer 20--30 kpc regions are shown in the right panel of Fig.~\ref{fig:sb_oii} for the matched samples. In the inner region, the \oii\ SB increases by a factor of $\approx$8 from \mstar\ = $10^{6-8}$\,\msun\ to $10^{10-12}$\,\msun\ for the redshift matched samples, and by a factor of $\approx$2 when matched additionally in SFR. In the outer region, the increasing trend with stellar mass is consistent within the large uncertainties for both the matched samples. Therefore, the dependence of the \oii\ emission on stellar mass is likely driven by a combination of dependence on star formation activity and other physical conditions in the gas. We note that in the above analysis, as mentioned in Sect.~\ref{sec:obs}, we corrected the fluxes for dust before stacking using the stellar extinction obtained from SED fitting, assuming that the nebular and stellar extinction follow each other.

\subsection{Line ratios around galaxies}
\label{sec:results_ratios}

\begin{figure*}
    \centering
    \includegraphics[width=0.32\textwidth]{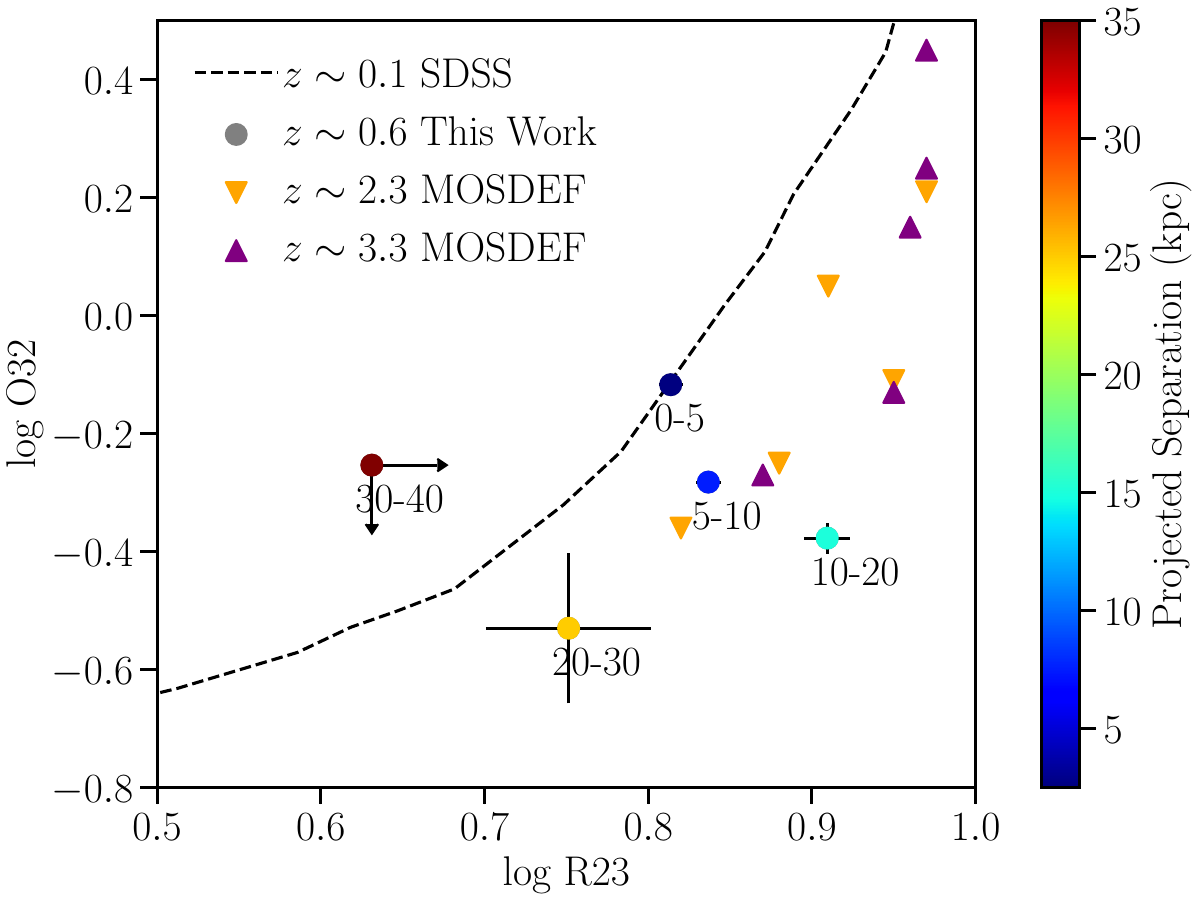}
    \includegraphics[width=0.32\textwidth]{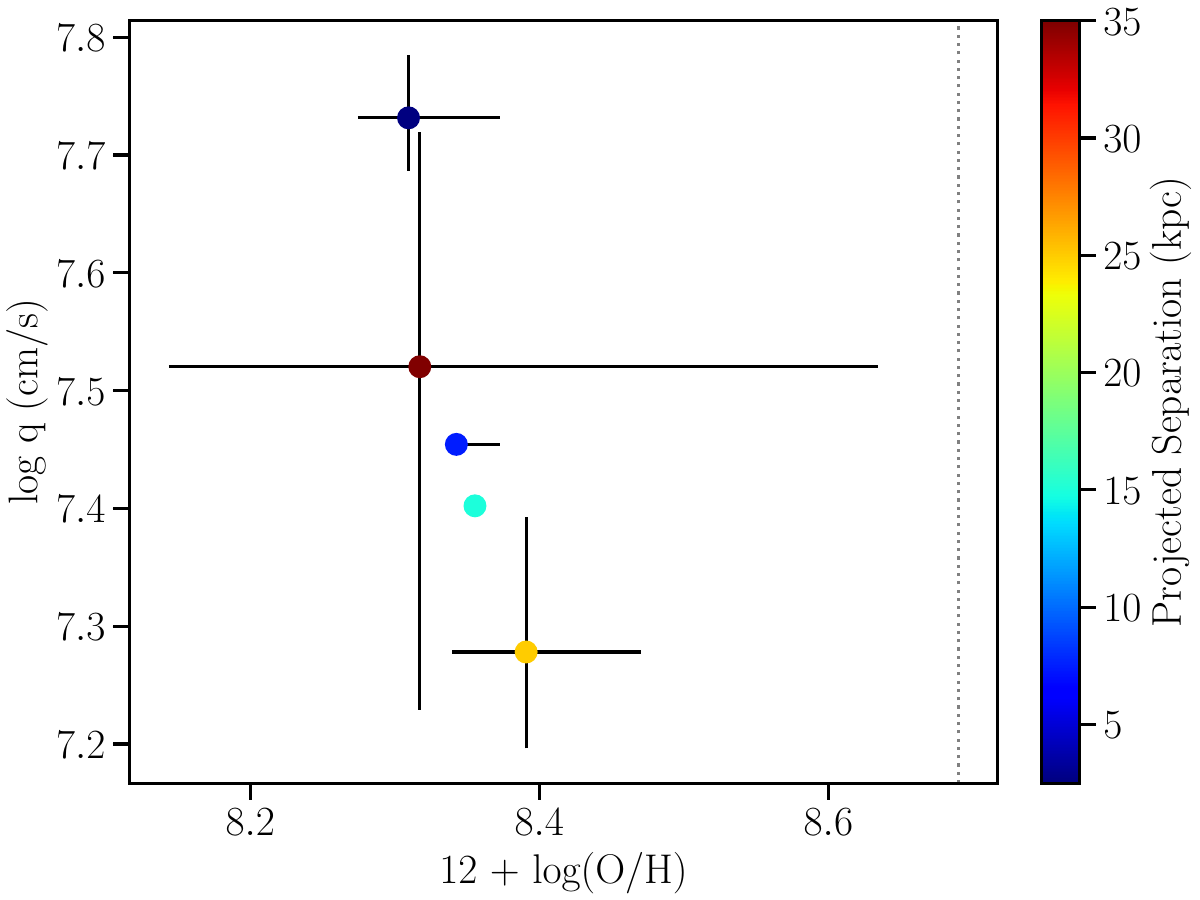}
    \includegraphics[width=0.32\textwidth]{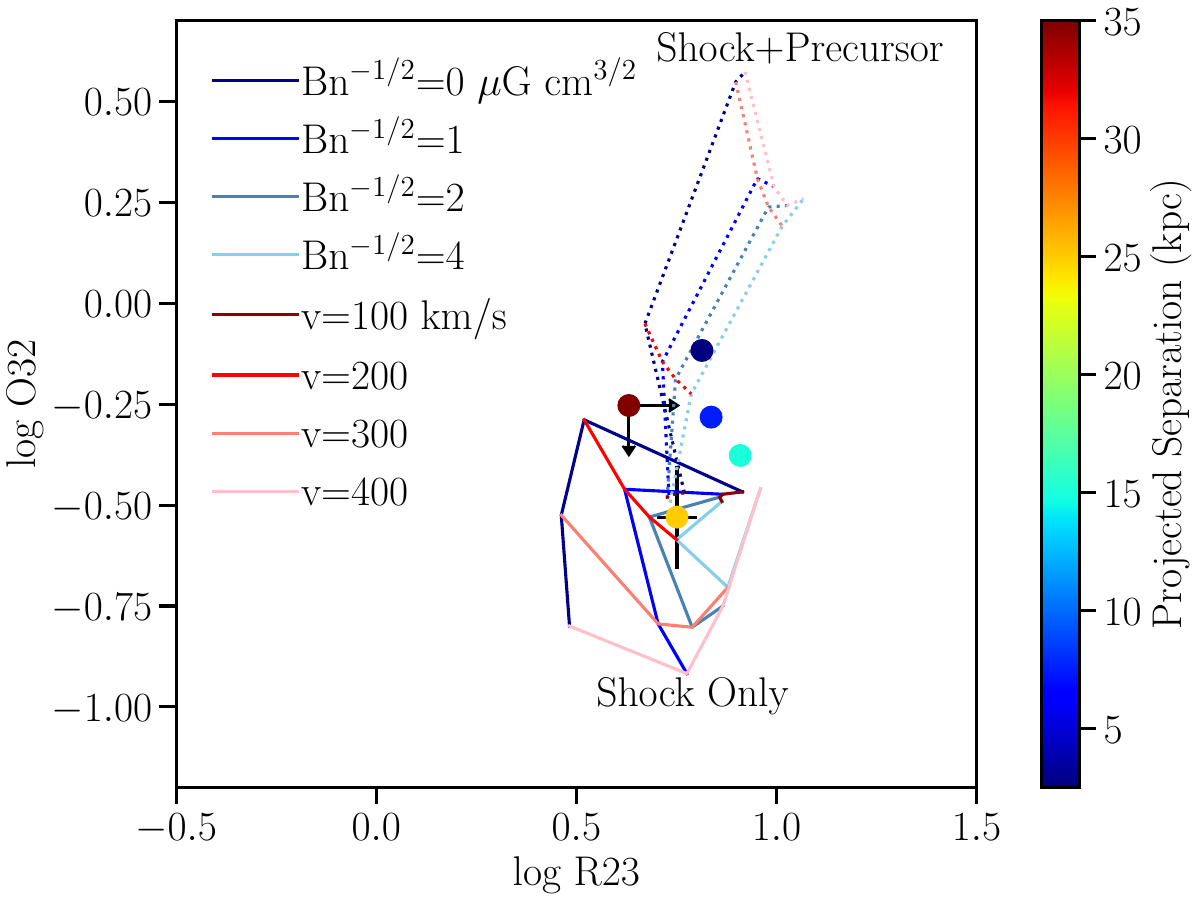}
    \caption{Stacked line ratios around galaxies. Left: O32 versus R23 diagram. The line ratios estimated from the stacked spectra (Fig.~\ref{fig:spectra}) are shown as circles that are color-coded by the average projected separation from the galaxy center. The annular bins (in kpc) corresponding to the ratios are also labeled next to each circle. Arrows denote $2\sigma$ limits on the line ratios. The median value of the line ratios of $z\approx0.1$ galaxies from SDSS is shown as the dashed black line. Line ratios from composite spectra of galaxies in the MOSDEF survey \citep{Sanders2021} are shown as downward orange triangles for the $z\approx2.3$ sample and as purple upward triangles for the $z\approx3.3$ sample.
    Center: Ionization parameter versus metallicity (symbols are same as in the left panel) estimated from the emission line fluxes in the stacked spectra and photoionization models using the code {\sc izi}. The solar metallicity is demarcated by the dotted line.
    Right: O23 versus R23 diagram of the stacked sample (symbols are same as in the left panel) compared with results from the ``shock only'' (solid lines) and ``shock+precursor'' models (dotted lines) from \citet{Allen2008}. Shock models are shown for several magnetic parameters and shock velocities, as denoted in the plot.
    }
    \label{fig:ratios}
\end{figure*}

Having detected emission from multiple lines in the stacked cube, we next investigated the average physical conditions in the ionized gas around $z\approx0.6$ galaxies using different line ratios. In particular, the O32 line ratio, defined as \oiii$\lambda$5008/\oii$\lambda\lambda$3727,3729, is an indicator of the degree of ionization of the gas, while the R23 line ratio, defined as (\oiii$\lambda$4960,5008+\oii$\lambda$3727,3939)/\hb, is sensitive to the gas-phase metallicity. The O32 versus R23 diagram has been used to study the ionization and metallicity in the ISM using integrated spectra of galaxies. Star-forming galaxies generally follow a trend in the O32--R23 diagram, going from low R23 and O32 values at high metallicity to high R23 and O32 values at low metallicity, although R23 begins to decrease at very low metallicity \citep[$\lesssim0.1Z_\odot$;][]{Kewley2019}. We note that according to the mass-metallicity relation, the median stellar mass of the sample is approximately at the turn-over metallicity of R23.

Thanks to stacking of the MUSE 3D data, here we were able to spatially resolve the above line ratios and study how they vary on average around a statistical sample of star-forming galaxies. We note that physical conditions such as metallicity can vary with stellar mass and redshift \citep{Tremonti2004,Maiolino2019}. Therefore, we created similar stacks in bins of stellar mass and redshift below and above the median values, and also by weighting with the stellar mass, to confirm that the trends with physical separation we find below remain similar over the stellar mass and redshift range probed here. 

To compute the line ratios, we integrated the flux in the spectra extracted from the stacked cube over the velocity window marked in Fig.~\ref{fig:spectra}. For the \oii\ doublet, we considered the total flux of both the lines. The error in the integrated flux was obtained by propagating the error in the spectra. If the integrated flux was detected at less than 2$\sigma$, we considered it an upper limit. The left panel of Fig.~\ref{fig:ratios} shows the O32 versus R23 line ratios estimated from the stacked spectra in different annuli (Fig.~\ref{fig:spectra}). The O32 ratio is found to decrease going from the central 0--5 kpc bin (log O32 $\approx-0.1$) to the outer 20--30 kpc bin (log O32 $\approx-0.5$). The R23 ratio is found to increase from the central 0--5 kpc bin (log R23 $\approx$0.8) to the 10--20 kpc bin (log R23 $\approx$0.9), and then decrease in the 20--30 kpc bin (log R23 $\approx$0.75).

For comparison, we also plot the O32 versus R23 ratios from integrated spectra of lower- and higher-redshift galaxies. The median value of the ratios in the sample of $z\approx0.1$ galaxies from the Sloan Digital Sky Survey \citep[SDSS;][]{York2000} are obtained using the Max-Planck-Institute for Astrophysics -- Johns Hopkins University (MPA--JHU) catalog\footnote{\url{https://wwwmpa.mpa-garching.mpg.de/SDSS/DR7/}}. The line ratios at $z\approx2.3$ and $z\approx3.3$ are obtained from the composite spectra of star-forming galaxies in different stellar mass bins ($\approx2\times10^9 - 4\times10^{10}$\,\msun) in the MOSFIRE Deep Evolution Field (MOSDEF) survey \citep{Sanders2021}, with the line ratios decreasing with increasing stellar mass. The stacked line ratios in the central region, within 20 kpc, are similar to the average line ratios found in the ISM of $z\approx2-3$ galaxies with stellar mass, log \mstar\ $\approx10.2-10.6$\,\msun. Furthermore, the stacked line ratios in the central region are consistent with the trend of R23 increasing at a fixed O32, and O32 decreasing at a fixed R23, from $z\approx0$ to $z\approx3$. On the other hand, the stacked line ratios at 20--30 kpc are more similar to those found in the ISM of local SDSS galaxies. This indicates a transition in the physical conditions or the ionization mechanism of the warm ionized gas from the inner to the outer regions (i.e., from the ISM to the disk-halo interface and the CGM).

To investigate this further, we compared our results with photoionization models. Strong line ratios -- calibrated either theoretically via comparison with photoionization models or empirically via comparison with estimates from direct method in sources when possible -- have been used extensively to infer the ionization parameter and chemical abundance of the photoionized gas in the ISM \citep{Maiolino2019,Kewley2019}. However, the O32 ratio, typically used as an estimator of the ionization parameter, is influenced by the gas-phase metallicity and gas pressure. Similarly, the R23 ratio, which is commonly used to determine chemical abundances, is sensitive to the ionization parameter and gas pressure, and is moreover double-valued, having both a low and a high abundance branch. On the other hand, Bayesian methods that use theoretical photoionization models to fit emission line fluxes can be used to simultaneously derive the ionization parameter (q) and gas-phase metallicity (Z), although they likely still suffer from some of the same limitations as the individual line ratio calibrations. 

Here, we used the code {\sc izi} \citep{Blanc2015,Mingozzi2020} to simultaneously infer q and Z in a Bayesian method by comparing the emission line fluxes in the stacked spectra with photoionization models. We adopted the models of \citet{Levesque2010} that were calculated using the high mass-loss tracks, a constant star formation history, a Salpeter IMF with a 100\,\msun\ upper cutoff, an electron density of 100\,\cc, and an age of 6 Myr. We verified that the radial trends of q and Z remain similar using the \citet{Levesque2010} models with different parameters or different photoionization models \citep[e.g.,][]{Kewley2001,Dopita2013}. We note that the photoionization models assume that the density of the emitting gas is constant across the scales of interest here, which may not be valid. The central panel of Fig.~\ref{fig:ratios} shows the q and Z values estimated by {\sc izi}, which also takes into account upper limits on the line fluxes. The ionization parameter is found to decrease from log q (cm~s$^{-1}$) $\approx$7.7 to $\approx$7.3 on going outward from the central 0--5 kpc region to the outer 20--30 kpc region, most likely due to a weaker radiation field in the outskirts of galaxies. While there appears to be a transition in the line ratios from the inner 20 kpc to the outer 20--30 kpc region in the O32--R23 diagram, there is no significant variation in the gas-phase metallicity between the inner and outer regions based on photoionization models, with 12 + log(O/H)\footnote{Solar metallicity, 12 + log(O/H) = 8.69 \citep{Asplund2009}} going from $\approx$8.3 at $\le$5 kpc to $\approx$8.4 at 20--30 kpc. Using the empirical calibrations of \citet{Curti2017}, we get a similar trend of metallicity, with 12 + log(O/H) $\approx$8.4 at $\le$5 kpc and $\approx$8.5 at 20--30 kpc.

The above analysis assumes that the warm ionized gas is photoionized. While this assumption is likely valid in the ISM, it may not necessarily hold in the extended disk or CGM, where shocks and turbulence could also ionize the gas. To investigate alternate ionization mechanisms, we compared the line ratios from the stacked spectra with shock models. The right panel of Fig.~\ref{fig:ratios} compares the O32 and R23 line ratios from the stacked spectra with shock models assuming a Large Magellanic Cloud abundance and a pre-shock density of 1\,\cc\ for several magnetic parameters and shock velocities from \citet{Allen2008}. The ``shock only'' models include contribution from radiative shocks only, while the ``shock+precursor'' models include contribution from the H\,{\sc ii} region ahead of the shock front as well. The line ratios in the inner 5 kpc region are consistent with the ``shock+precursor'' models with magnetic parameter, Bn$^{-1/2}$ = 2--4 $\mu$G cm$^{3/2}$, and velocity 200\,\kms. This suggests that photoionization is the dominant ionization mechanism in the central region. On the other hand, at 20--30 kpc, the line ratios are consistent with the ``shock only'' models with magnetic parameter, Bn$^{-1/2}$ = 2--4 $\mu$G cm$^{3/2}$, and velocity 200\,\kms. This indicates that outer regions could be affected by radiative shocks arising due to gas flows and interactions, although the observed velocity dispersions of the emission lines are less than the required shock velocities.

\section{Discussion and summary} 
\label{sec:discussion}

We have presented the average line emission around a statistical sample of 560 galaxies at $z\approx0.25-0.85$ based on stacking of the MUSE 3D data of two large surveys, MAGG and MUDF. The redshift range was selected to have simultaneous coverage of the \oii\ $\lambda\lambda$3727,3729, \hb\ $\lambda$4863, \oiii\ $\lambda$4960, and $\lambda$5008 emission lines. The average \oii\ line emission is detected out to $\approx$40 kpc, and the average \oiii\ and \hb\ emission is detected out to $\approx$30 kpc. For comparison, the average stellar continuum emission is detected out to $\approx$20 kpc. Therefore, the extended emission likely probes the disk-halo interface or the transition region between the ISM and the CGM. 

Combining our sample with that of \citet{Dutta2023}, who conducted a similar stacking analysis at higher redshifts, we find that the \oii\ emission increases independently with both redshift over $z\approx$ 0.4--1.3 and with stellar mass over \mstar\ $\approx10^{6-12}$\,\msun. Based on a control analysis, we find that the enhancement in the average \oii\ SB in the inner 20 kpc region tracing the ISM is driven to a large extent by the higher SFRs at higher redshifts and stellar masses. On the other hand, additional factors, such as the evolution of physical conditions in the warm ionized gas, likely contribute to the observed enhancement in the \oii\ SB in the outer 20--30 kpc region with redshift and stellar mass.

To investigate the average physical conditions in the extended disks or the disk-halo interface of galaxies, we analyzed the emission line ratios in the stacked spectra. Both the O32 and R23 line ratios are found to decrease from the center out to 30 kpc. The line ratios at $<$20 kpc occupy a similar region in the O32--R23 diagram as those obtained from integrated spectra probing the ISM of $z\approx2-3$ galaxies. On the other hand, the line ratios at $>$20 kpc are shifted in the O32--R23 diagram, which indicates a transition in the physical conditions and/or ionization mechanisms of the warm ionized gas from the ISM to the CGM.

We simultaneously estimated the ionization parameter and gas-phase metallicity under the assumption that the gas is photoionized. For this purpose, we used the Bayesian code {\sc izi} to compare the emission line fluxes in the stacked spectra with the photoionization models of \citet{Levesque2010}. The ionization parameter exhibits a declining trend from the central region out to 30 kpc. This is consistent with a picture in which the number of \hii\ regions decreases with distance from galaxy center, resulting in a weaker radiation field in the outskirts. On the other hand, the metallicity does not show any significant trend, remaining more or less similar up to 30 kpc. As discussed in Sect.~\ref{sec:intro}, several mechanisms acting simultaneously are likely influencing the distribution of metals within and around galaxies. For example, the mixing of metals due to the balance between the inflow of metal-poor gas, outflow of metal-rich gas, and gas transfer due to mergers could lead to a flat metallicity gradient. Several studies have found that the metallicity gradients in the ISM of $z\gtrsim0.5$ galaxies are generally flat with a large scatter \citep[e.g.,][see however \citealt{Moller2020}]{Wuyts2016,Carton2018,Curti2020,Wang2020,Simons2021,Venturi2024}. Studies of the CGM metallicity have also indicated  that the CGM is complex, with gas outflows, accretion, and recycling all at play \citep[e.g.,][]{Lehner2019,Pointon2019,Guo2020}. Extending such studies into the disk-halo interface, we find that the average metallicity gradient is flat ($\approx$0.003 dex kpc$^{-1}$) out to 30 kpc, indicating an efficient mixing of metals at these larger scales. 

We point out that other ionization mechanisms in addition to photoionization by H\,{\sc ii} regions might be altering the observed line ratios and metallicity measurements in the outer regions. For example, the gas could be ionized by radiative shocks arising due to stellar outflows, supernovae, gas accretion from the CGM, or interactions \citep[e.g.,][]{Fossati2016,Epinat2018,Chen2019,Franchetto2021}. By comparing the line ratios from the stacked spectra with those from the shock models of \citet{Allen2008}, we find that the line ratios at 20--30 kpc can be explained by shock velocities of $\approx$200\,\kms. However, it may be difficult to reconcile the observed narrow line widths ($\lesssim100$\,\kms) with the required shock velocities, although slow shock models with a velocity dispersion of $\approx$100\,\kms\ have been found to produce line ratios consistent with active galactic nucleus-like excitation due to shocks \citep{Rich2011}. In addition, there could be a contribution from the diffuse ionized gas to the line emission in the outer regions, leading to line ratios significantly different from those produced by H\,{\sc ii} regions \citep[e.g.,][]{Zhang2017,Vale2019,Mannucci2021}. The  diffuse ionized gas could be ionized by photons leaking out of H\,{\sc ii} regions, hot evolved stars, and shocks. In the future, additional multiwavelength observations such as NIR observations of \ha, \nii,\ and \sii\ emission lines would facilitate distinguishing between different ionization mechanisms.

We note that we stacked the emission of galaxies at random disk orientations, washing out any azimuthally dependent signatures of metallicity gradients if present, such as those due to biconical outflows of metal-rich gas from the center of galaxies or the accretion of metal-poor gas along the plane of the galaxy disks \citep[e.g.,][]{Bouche2013,Peroux2020b}. In \citet{Dutta2023}, the average \oii\ emission at $z\approx1$ was found to be $\approx$30\% more extended along the major axis than along the minor axis, while the average \mgii\ emission at $z\approx1$ was found to be enhanced up to 10 kpc along the minor axis in the stacking analysis of \citet{Guo2023}. In the current sample, the number of galaxies with orientation measurements from high-spatial-resolution HST imaging is not sufficiently large to probe extended emission from multiple lines in stacking. 

Nevertheless, the average metallicity gradient in a statistical sample can still place useful constraints on galaxy formation models. For example, comparing different feedback prescriptions in cosmological hydrodynamical simulations, \citet{Gibson2013} found that an enhanced feedback model gives rise to flat metallicity gradients in galaxies. Recently, using EAGLE simulations, \citet{Tissera2022} found that the median metallicity gradient is close to zero at all redshifts, with mergers and gas accretion both influencing the gradients, while \citet{Hemler2021} found predominantly negative metallicity gradients at all redshifts using TNG-50 simulations.\ This shows that the gradient can be a useful tool for discriminating between different feedback models. The flat metallicity gradient ($\approx$0.003 dex kpc$^{-1}$) out to 30 kpc found in this work supports a model in which feedback and interactions regulate the gradients. Future detections of extended emission in multiple lines from the gas around a larger sample of individual galaxies, as well as an extension of the stacking analysis to higher redshifts using IFU data from the JWST and the Extremely Large Telescope, are expected to constrain galaxy formation models even further by characterizing the radial gradients of the physical, chemical, and ionization conditions of gas from the ISM to the CGM.  

\begin{acknowledgements}

We thank the anonymous referee for their useful comments.

RD thanks Palle M\o{}ller for helpful comments.

This work has been supported by Fondazione Cariplo, grant No 2018-2329.

This work is based on observations collected at the European Organisation for Astronomical Research in the Southern Hemisphere under ESO programme IDs 197.A-0384 and 1100.A-0528. 

Based on observations with the NASA/ESA \textit{Hubble} Space Telescope obtained, from the Data Archive at the Space Telescope Science Institute, which is operated by the Association of Universities for Research in Astronomy, Incorporated, under NASA contract NAS 5-26555. Support for Program numbers 15637, and 15968 were provided through grants from the STScI under NASA contract NAS 5-26555.

This work used the DiRAC Data Centric system at Durham University, operated by the Institute for Computational Cosmology on behalf of the STFC DiRAC HPC Facility (\url{www.dirac.ac.uk}). This equipment was funded by BIS National E-infrastructure capital grant ST/K00042X/1, STFC capital grants ST/H008519/1 and ST/K00087X/1, STFC DiRAC Operations grant ST/K003267/1 and Durham University. DiRAC is part of the National E-Infrastructure.

\end{acknowledgements}

%
%

\bibliographystyle{aa}
\bibliography{paper}

\end{document}